\documentclass[letterpaper, 10 pt, conference]{ieeeconf}
\IEEEoverridecommandlockouts
\usepackage{cite}
\usepackage{amsmath,amssymb,amsfonts}
\usepackage{algorithmic}
\usepackage{graphicx}
\usepackage{textcomp}
\usepackage{xcolor}
\usepackage[hidelinks]{hyperref}
\usepackage{tabularx}

\newcommand{\orcid}[1]{\href{https://orcid.org/#1}{\includegraphics[width=8pt]{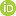}}}

\def\BibTeX{{\rm B\kern-.05em{\sc i\kern-.025em b}\kern-.08em
    T\kern-.1667em\lower.7ex\hbox{E}\kern-.125emX}}

\title{ \LARGE \bf Context Informed Incremental Learning Improves Myoelectric Control Performance in Virtual Reality Object Manipulation Tasks
}

\author{
    Gabriel~Gagné$^{1}$~\orcid{0009-0007-0958-6799},
    Anisha~Azad$^{2}$~\orcid{0009-0004-2612-3989},
    Thomas~Labbé$^{1}$~\orcid{0009-0003-1766-4022},
    Evan~Campbell$^{3}$~\orcid{0000-0001-5399-4318}, 
    Xavier~Isabel$^{1}$~\orcid{0009-0007-1354-5578},\\
    Erik~Scheme$^{3}$~\orcid{0000-0002-4421-1016},
    Ulysse~Côté-Allard$^{4, \dagger}$~\orcid{0000-0003-3241-8404},
    Benoit~Gosselin$^{1, \dagger}$~\orcid{0000-0003-1473-3451}
    \thanks{This work was supported in part by the Natural Sciences and Engineering Research Council of Canada and in part by Mitacs.}
    \thanks{$^{1}$G. Gagné, T. Labbé, X. Isabel and B. Gosselin are with the Department of Electrical and Computer Engineering, Laval University, Québec, QC, Canada.}
    \thanks{$^{2}$A. Azad is with the School of Informatics, University of Edinburgh, Newington, Edinburgh, United Kingdom.}
    \thanks{$^{3}$E. Campbell and E. Scheme are with the Department of Electrical and Computer Engineering, University of New Brunswick, Fredericton, NB, Canada.}
    \thanks{$^{4}$U. Côté-Allard is with the Department of Technology Systems, University of Oslo, Oslo, Norway.}
    \thanks{\small$^\dagger$ Shared last authorship.}
}

\begin{document}

\maketitle

\thispagestyle{empty}
\pagestyle{empty}

\begin{abstract}

Electromyography (EMG)-based gesture recognition is a promising approach for designing intuitive human-computer interfaces. However, while these systems typically perform well in controlled laboratory settings, their usability in real-world applications is compromised by declining performance during real-time control. This decline is largely due to goal-directed behaviors that are not captured in static, offline scenarios. To address this issue, we use \textit{Context Informed Incremental Learning} (CIIL) - marking its first deployment in an object-manipulation scenario - to continuously adapt the classifier using contextual cues. Nine participants without upper limb differences completed a functional task in a virtual reality (VR) environment involving transporting objects with life-like grips. We compared two scenarios: one where the classifier was adapted in real-time using contextual information, and the other using a traditional open-loop approach without adaptation. The CIIL-based approach not only enhanced task success rates and efficiency, but also reduced the perceived workload by 7.1\%, despite causing a 5.8\% reduction in offline classification accuracy. This study highlights the potential of real-time contextualized adaptation to enhance user experience and usability of EMG-based systems for practical, goal-oriented applications, crucial elements towards their long-term adoption. The source code for this study is available at: \url{https://github.com/BiomedicalITS/ciil-emg-vr}.

\end{abstract}


\section{Introduction}


Gesture recognition is a promising approach for developing advanced and intuitive human-computer interfaces (HCI)~\cite{mitra_gesture_2007}. With advances in wearable technology and machine learning, electromyography (EMG) has emerged as a key HCI modality, enabling applications such as operating prosthetic devices, sign language interpretation or robotics control~\cite{scheme_electromyogram_2011, sign_language_EMG, cote-allard_deep_2019}. By analyzing muscle electrical activity during gestures, EMG systems can predict hand and wrist movements in real-time~\cite{phinyomark_surface_2020}. Compared to other modalities like computer vision~\cite{vision-based_gesture_recognition}, EMG offers several advantages: it is robust to lighting and field-of-view challenges, it sidesteps privacy concerns associated with environmental recording, and is uniquely suited for operating prosthetic devices via residual muscle activity.


Despite the advantages of myoelectric control systems, their practical use is hindered by a decline in performance in real-world settings, affecting their long-term usability~\cite{cote-allard_unsupervised_2020}. Notably, EMG signals recorded during goal-directed tasks often differ significantly from those captured during controlled, screen-guided training sessions~\cite{campbell_screen_2024}. This discrepancy is compounded by confounding factors such as variations in electrode placement, muscle fatigue, and user-specific differences, all of which limit the reliability of traditional supervised calibration approaches~\cite{campbell_current_2020}. Additionally, variability in hardware, gesture sets, and control schemes restricts the effective use of large datasets, necessitating user and session-specific models~\cite{jiang_bio-robotics_2023, sultana_systematic_2023}.


To address these challenges and reduce the reliance on regular dedicated calibration sessions, migrating to models that continuously adapt to the user’s behavior during the control task is critical. Context-informed Incremental Learning (CIIL) offers a promising approach by leveraging contextual knowledge to generate pseudo-labels during task performance~\cite{campbell2024context, campbell_screen_2024}. These pseudo-labels allow for real-time, incremental adaptation of the classification algorithm, enabling the system to adjust dynamically to diverse and evolving conditions. This is particularly beneficial for real-time tasks, where participants are focused on achieving a goal rather than deliberately performing isolated gestures for the system’s benefit, as is typically the case during screen-guided training (SGT).

Previous research on CIIL has primarily used 2D video games to evaluate the feasibility of these methods in the context of EMG-based human-input devices~\cite{campbell2024context, campbell_screen_2024}. These studies have demonstrated the potential of real-time adaptation but are limited by their testing environments, which lack the complexity and functional relevance of real-world scenarios, especially within the context of prosthetic control. In contrast to previous studies, this work evaluates the CIIL framework in a virtual reality (VR) environment, focusing on realistic grasping tasks. This setup allows for natural object manipulation, mirroring real-world prosthetic challenges, and providing a more comprehensive evaluation of CIIL in EMG-based systems.

This paper is organized as follows. Section~\ref{sec:methodology} details the hardware and software components of the system. Section~\ref{sec:results} presents the results, with a subsequent discussion in Section~\ref{sec:discussion}.

\section{Methodology}
\label{sec:methodology}

\subsection{Experiment overview}

The experiment was divided into several stages (see Table~\ref{tab:procedure}). First, Steps 1-2 were used to generate and evaluate a baseline model, which served as the foundation for the subsequent stages. Next, Steps 3-5 were repeated twice to create two distinct trials: one with live adaptation (CIIL trial) and one with no adaptation (NA trial). The trial order was randomized to avoid a learning effect. After each trial, the system was evaluated both subjectively—using a NASA Task Load Index (NASA-TLX) survey~\cite{hart_nasa_1986, kosch_survey_2023}—and objectively, based on classification accuracy as well as online task-specific metrics like completion rate and time spent on each object.

\begin{table}
\centering
\caption{Experiment procedure}
    \begin{tabularx}{0.95\columnwidth}{ p{2.7cm} X }
    \hline
    \textbf{Step} & \textbf{Details} \\ \hline
    1. Offline training & Screen-guided procedure: 3 repetitions of 3 seconds. \\ 
    2. Initial testing & 2 screen-guided testing repetitions of 3 seconds. \\ 
    3. Online testing & In a VR environment, 6 objects had to be moved from one table to another in under 5 minutes. \\ 
    4. Subjective evaluation & The NASA-TLX workload assessment gives a subjective evaluation of the classifier. \\
    5. Post-task testing & 2 screen-guided testing repetitions of 3 seconds for an objective evaluation of the classifier. \\
    6. Repeat steps 3 to 5 & Repeat for the other trial. \\ \hline
    \end{tabularx}
\label{tab:procedure}
\end{table}

The source code for this study is available on GitHub\footnote{Source code: \url{https://github.com/BiomedicalITS/ciil-emg-vr}}. The software is distributed between two main processes: the Python machine learning server and the Unity virtual reality game. The server makes extensive use of the \textit{LibEMG} Python library~\cite{10214558}.

\subsection{Hardware \& Signal Processing}
\label{ssec:system_hardware}

\begin{figure}
    \centerline{\includegraphics[width=\columnwidth]{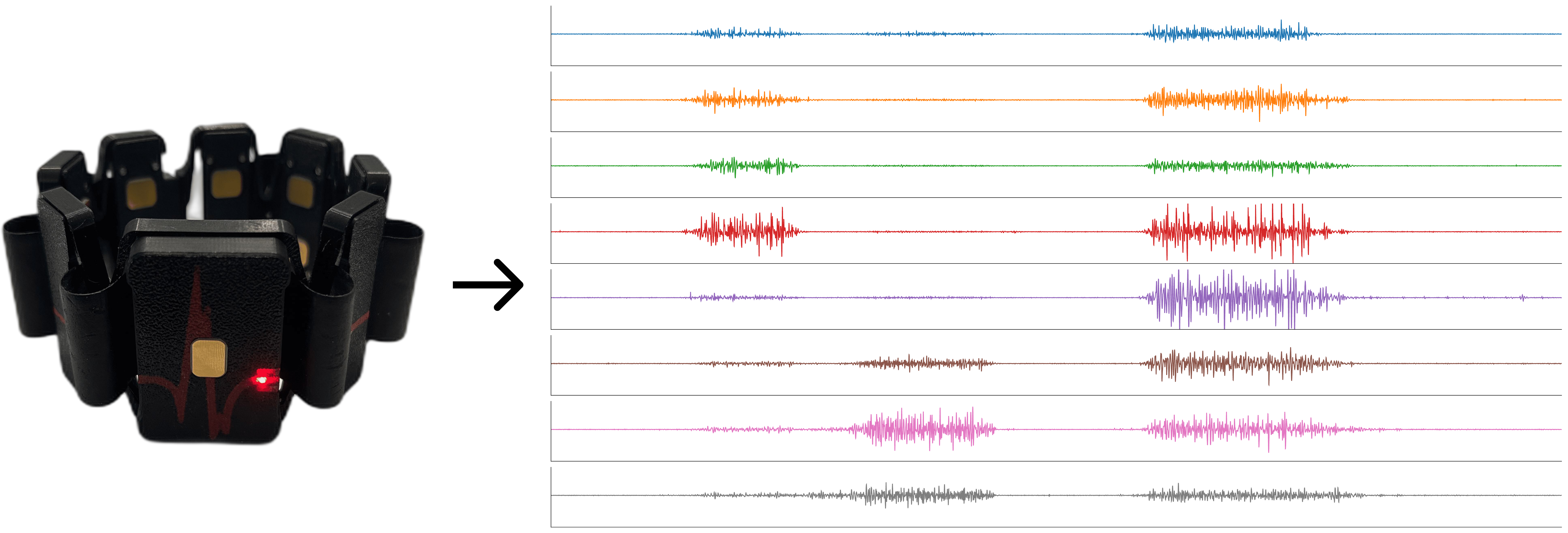}}
    \caption{The 8-channel (2000 Hz) SiFiBand from SiFi Labs Inc. used in this study (left), along with the corresponding raw EMG waveforms (right) recorded over 4 seconds during wrist flexion, wrist extension, and a power grip gesture.}
    \label{fig:bioarmband-example}
\end{figure}

Throughout the study, data were recorded using the SiFiBand by SiFi Labs Inc. The SiFiBand is a wireless wearable device that provides real-time 8-channel EMG, ECG, PPG, IMU, EDA, and skin temperature data. Within the scope of this work, only the 8-channel EMG was used, sampled at 2000~Hz. Fig.~\ref{fig:bioarmband-example} shows the armband and the corresponding EMG data recorded during various gestures.

The signal processing pipeline consisted of on-device 20-450~Hz band-pass filtering and power line noise removal~\cite{chen_review_2023}. The EMG data were accumulated into 200~ms long overlapping windows, with a stride of 50~ms. These windowing parameters ensured a good compromise between response latency and performance~\cite{optimal_windowsize_250ms}. Each window was used to extract the Time Domain Power Spectral Descriptors (TDPSD) feature set~\cite{al-timemy_improving_2016}. The resulting frames had dimensions of 6 features and 8 EMG channels.

\subsection{Classification Algorithm}

The gesture classifier was a one-dimensional convolutional neural network. Its architecture is detailed in Table~\ref{tab:cnn}. The geometry of the SiFiBand enables it to sense EMG signals around the entire forearm. Thus, the Conv1D layers were used to benefit from each gesture's specific spatial activation patterns, while the TDPSD feature set (with each feature used as an input channel) extracted information from the time and frequency domains, thereby maximizing the amount of information from which the model was able to learn.

\begin{table}[t!]
\centering
\caption{Convolutional model architecture}
    \begin{tabular}{ l l l l }
    \hline
    \textbf{Layer} & \textbf{Hyper-parameters} & \textbf{Add. layers} & \textbf{Activation} \\ \hline
    Conv1D & 32, 5x1 filt., 0-pad. & Batch norm. & Leaky ReLU \\ 
    Conv1D & 32, 3x1 filt., 0-pad. & Batch norm. & Leaky ReLU \\ 
    Dense & 256 out & Dropout (0.2) & Leaky ReLU \\ 
    Dense & 8 out & & \\ \hline
    \end{tabular}
\label{tab:cnn}
\end{table}

\subsection{Data Collection}

\begin{figure}
    \centering
    \includegraphics[width=\linewidth]{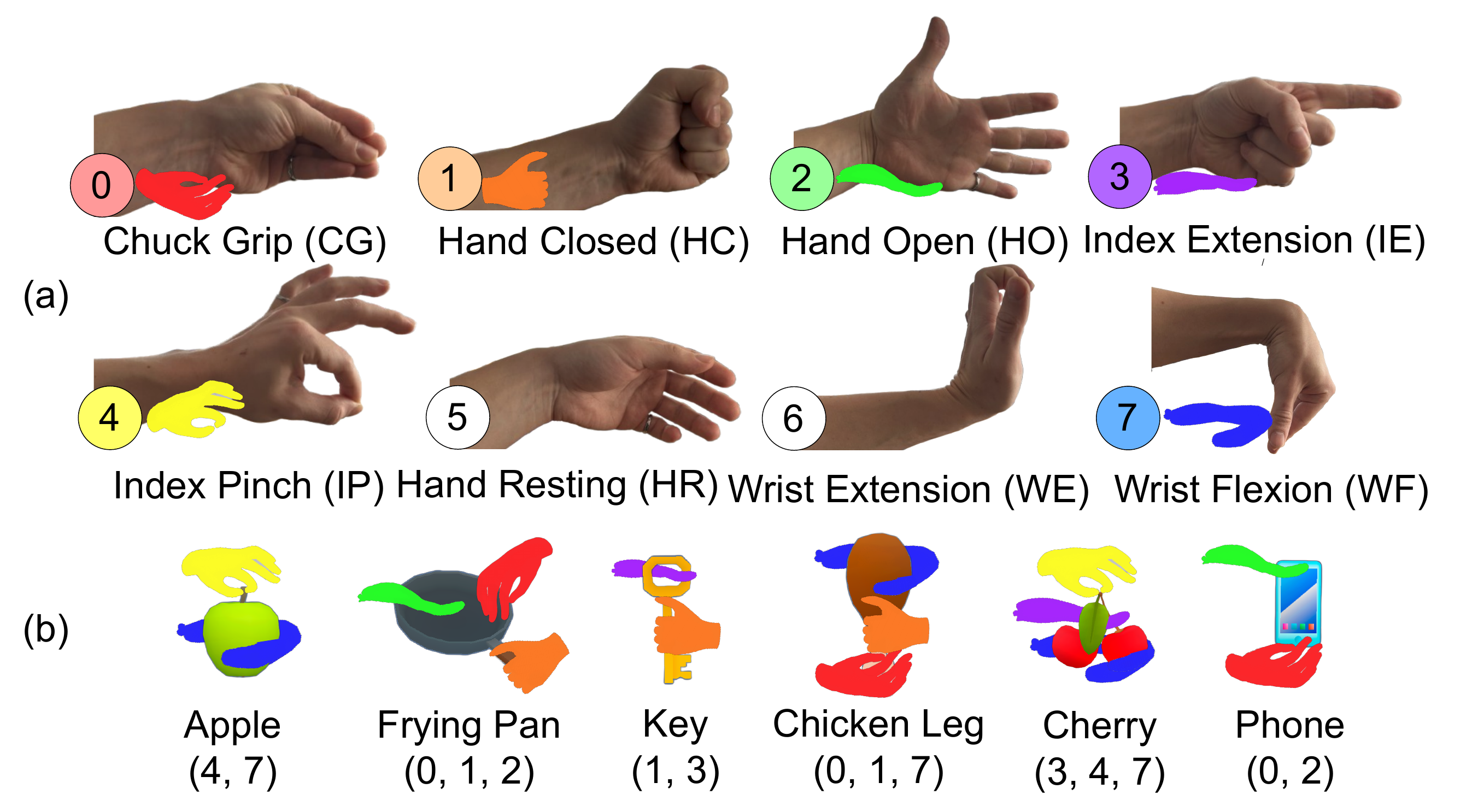}
    \caption{(a) Set of gestures included in the experiment and their associated VR game color and hand model. (b) Set of items to move from one table of another during the VR task and their corresponding valid grasp gestures.}
    \label{fig:gestures}
\end{figure}

In this study, data were collected both in an offline and online setting. All data were acquired with the SiFiBand, which was slid up until it fit snugly on the participant's forearm. The set of gestures used is illustrated in Fig.~\ref{fig:gestures} (a). Gestures were selected to align with those commonly used in day-to-day life, to ensure they are of functional interest. The numbered circles and hand models correspond to the mapping between class number and VR game hand pose. The HR and WE classes were not mapped to a VR pose, providing the opportunity to evaluate how the adapted classifier performed on classes which were not seen throughout online use. 

The data acquisition protocol was approved by the ethics committee for sectorial research in readaptation and social integration of the CIUSSS de la Capitale-Nationale (project 2023-2639). Informed consent was obtained from all participants, ensuring ethical compliance. This study includes data from nine participants without upper limb differences, aged between 21 and 57 years~(age: $32 \pm 13$, 4 women and 5 men), varying from first-time to experienced myoelectric users. The experiment lasted approximately 1 hour.

\subsubsection{Offline Training \& Evaluation}

Offline training started with a standard SGT protocol. Participants performed each of the eight gestures for 3 seconds, repeating each gesture five times, while only the isometric phase was recorded. The first three repetitions were used to create the training dataset, while the last two were used for initial testing. This procedure established a baseline SGT-based model for later stages. The same testing method was applied during the post-VR task evaluation, where participants completed two 3-second repetitions following each VR trial.

The model was trained for 15 epochs with the AdamW optimizer with $lr = 0.001$ and cross-entropy loss.

\subsubsection{Online Evaluation}

\begin{figure}
    \centerline{\includegraphics[width=\columnwidth]{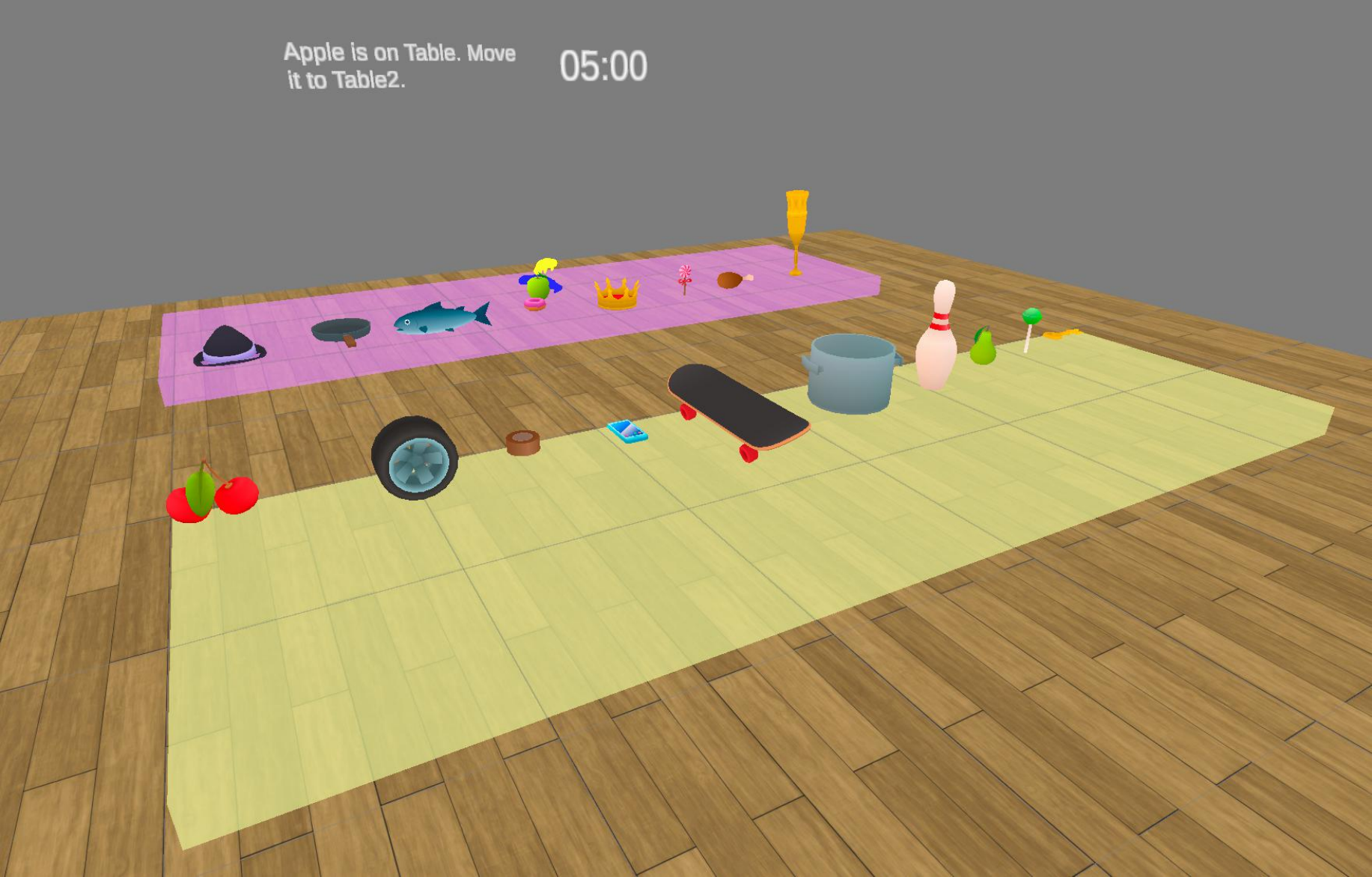}}
    \caption{Interactive VR environment of this study. Throughout the trials, instructions and a timer were displayed to guide subjects.}
    \label{fig:VRgame}
\end{figure}

The Unity VR environment, shown in Fig.~\ref{fig:VRgame}, was used to simulate a physical transport task of six objects between two tables\footnote{A demo video is available at: \url{https://youtu.be/m1aEv54v4n4}}. The experiment featured two trial types: one without adaptation (No Adaptation, NA) and the other with adaptation (CIIL trial). In each trial, subjects moved toward an object, began a grip by activating their forearm muscles with color-coded feedback from the classifier. When a valid grip was detected and the hand was within 10 cm of the object, the system registered a continuous hold. Then, participants could transport the object to the opposite table. Hand tracking was accomplished with the Meta Quest 2's native hand tracking. Finally, the object was deposited on the table by releasing the grasp. During these interactions, EMG signals, classification predictions, and environment context (i.e., spatio-temporal position of the hands and objects, grabbing status) were stored for online evaluation and for adaptation within the CIIL trial.

\subsubsection{Online Adaptation}

At each time step (i.e., game frame), the Unity backend tracked the spatial relationship between the participant’s hand and the nearest object to manage interactions and determine context suitability and available grasp options. When the hand entered the object’s grasping radius (10 cm), the classifier’s prediction was compared with the set of grasps associated with that object (Fig.~\ref{fig:gestures}b). If the prediction matched an allowed grasp, the context suitability for that window was labeled as \textit{positive}. Otherwise, it was labeled as \textit{negative}.

It is important to note that context suitability was not used to influence the classifier’s decision during real-time operation. Instead, it served solely to guide the incremental updates of the classifier, following recent CIIL methodologies~\cite{campbell_screen_2024}. Positive context suitability indicated a high likelihood that the classifier’s prediction was correct, while negative suitability suggested that the prediction was likely incorrect and that alternative grasp options should be considered. Accordingly, the adaptation dataset was built dynamically: for positive context windows, pseudo-labels were assigned based on the classifier’s prediction, while for negative context windows, pseudo-labels were generated by equally weighting all grasps associated with the object.

Every two seconds during the CIIL trial, accumulated data windows were used to retrain the model for one epoch, incorporating context-based pseudo-labels. Label spreading~\cite{zhu2002learning} was applied to refine the pseudo-labels, treating positive suitability samples as supervised (high contribution, resistant to relabeling) and negative suitability samples as unsupervised (low contribution, easily relabeled), with $\alpha=0.2$. This approach allowed the system to adapt continuously, incorporating subtle variations in user behavior.

\section{Results}
\label{sec:results}

\begin{figure*}[t]
    \centerline{\includegraphics[width=\textwidth]{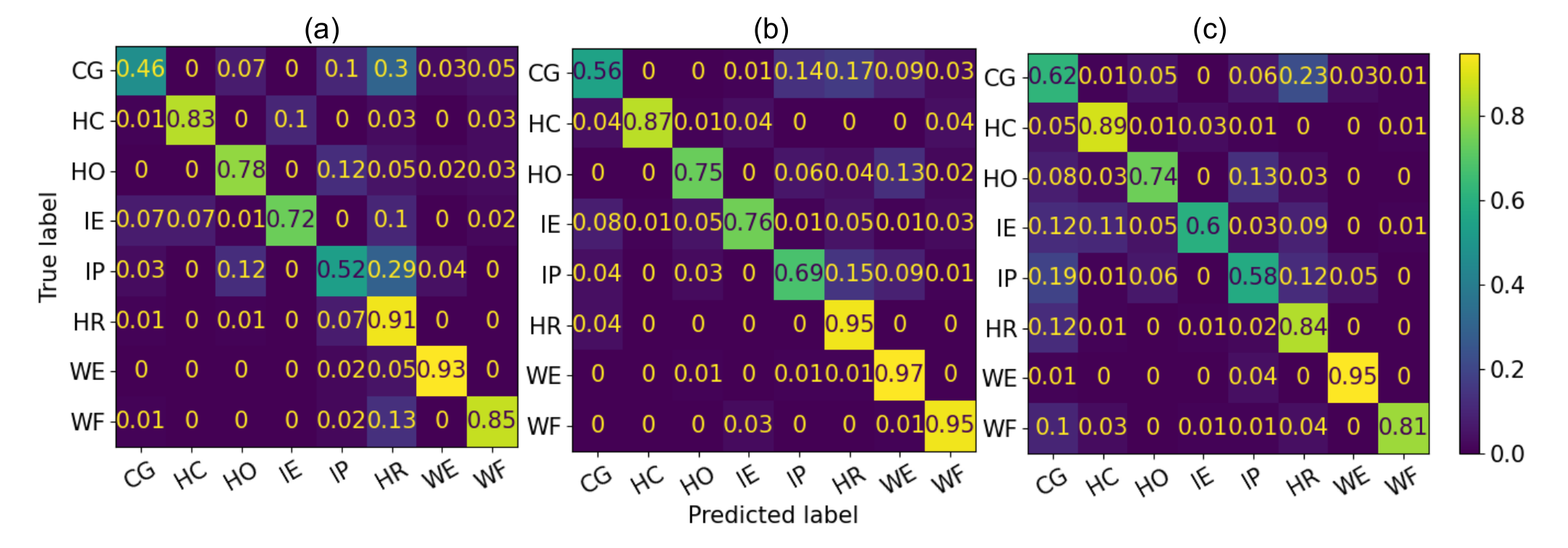}}
    \caption{Offline testing confusion matrices. (a) Initial test, (b) post-NA test and (c) post-CIIL test.}
    \label{fig:results-heatmap}
\end{figure*}

\begin{figure}[t]
    \centerline{\includegraphics[width=\linewidth]{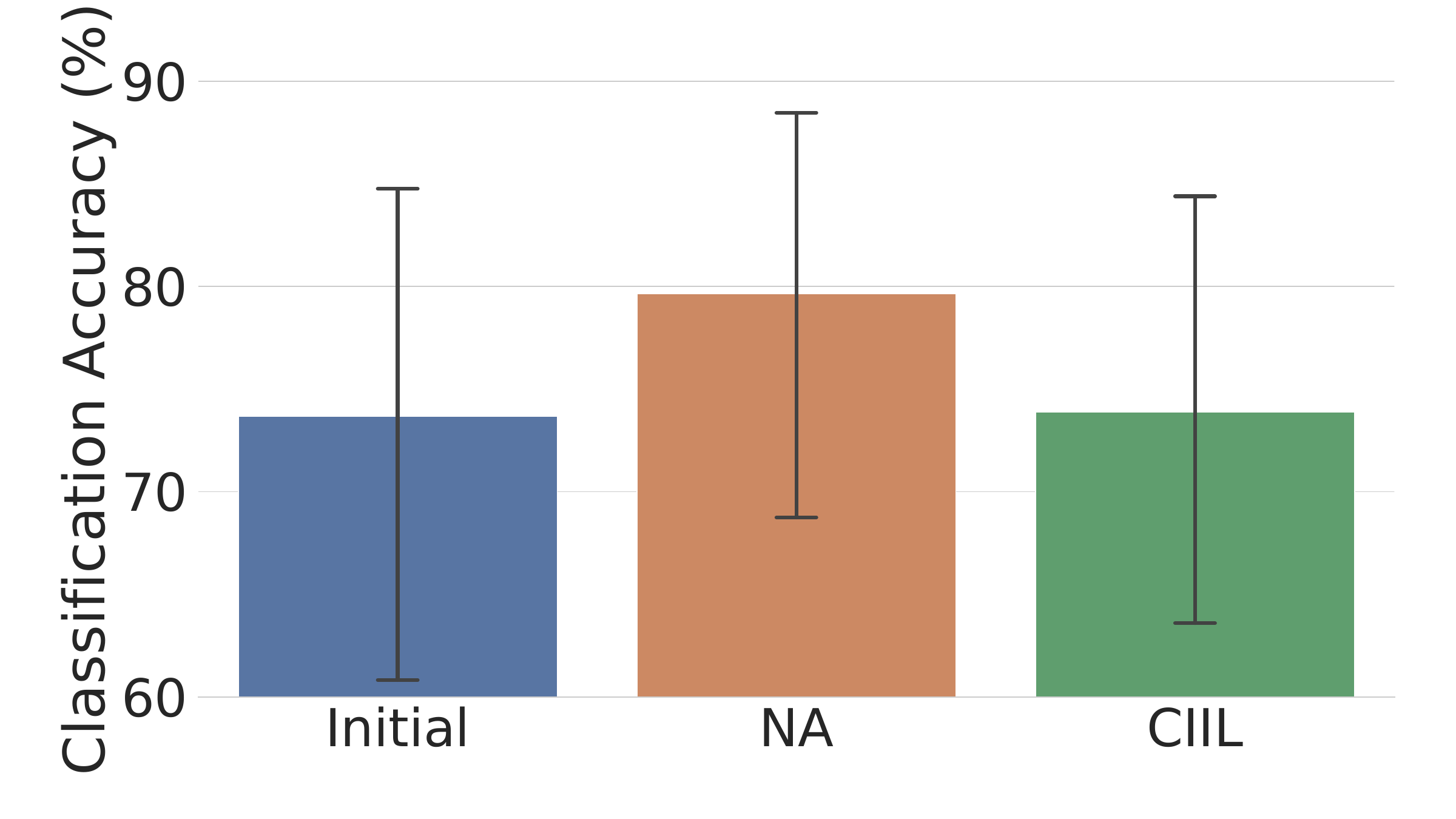}}
    \caption{Offline classification accuracy results of the Initial, post-NA and post-CIIL testing sessions.}
    \label{fig:results-boxplot}
\end{figure}

The offline evaluation results are presented in Fig.~\ref{fig:results-boxplot}. The classification accuracy obtained from screen-guided testing repetitions was $73.68 \pm 18.38$\%, $79.64 \pm 14.89$\% and $73.87 \pm 15.64$\% for the Initial, Post-NA and Post-CIIL cases, respectively. Fig.~\ref{fig:results-heatmap} shows the confusion matrices for all three offline test steps, averaged over all participants. 

\begin{figure}[t]
    \centerline{\includegraphics[width=\linewidth]{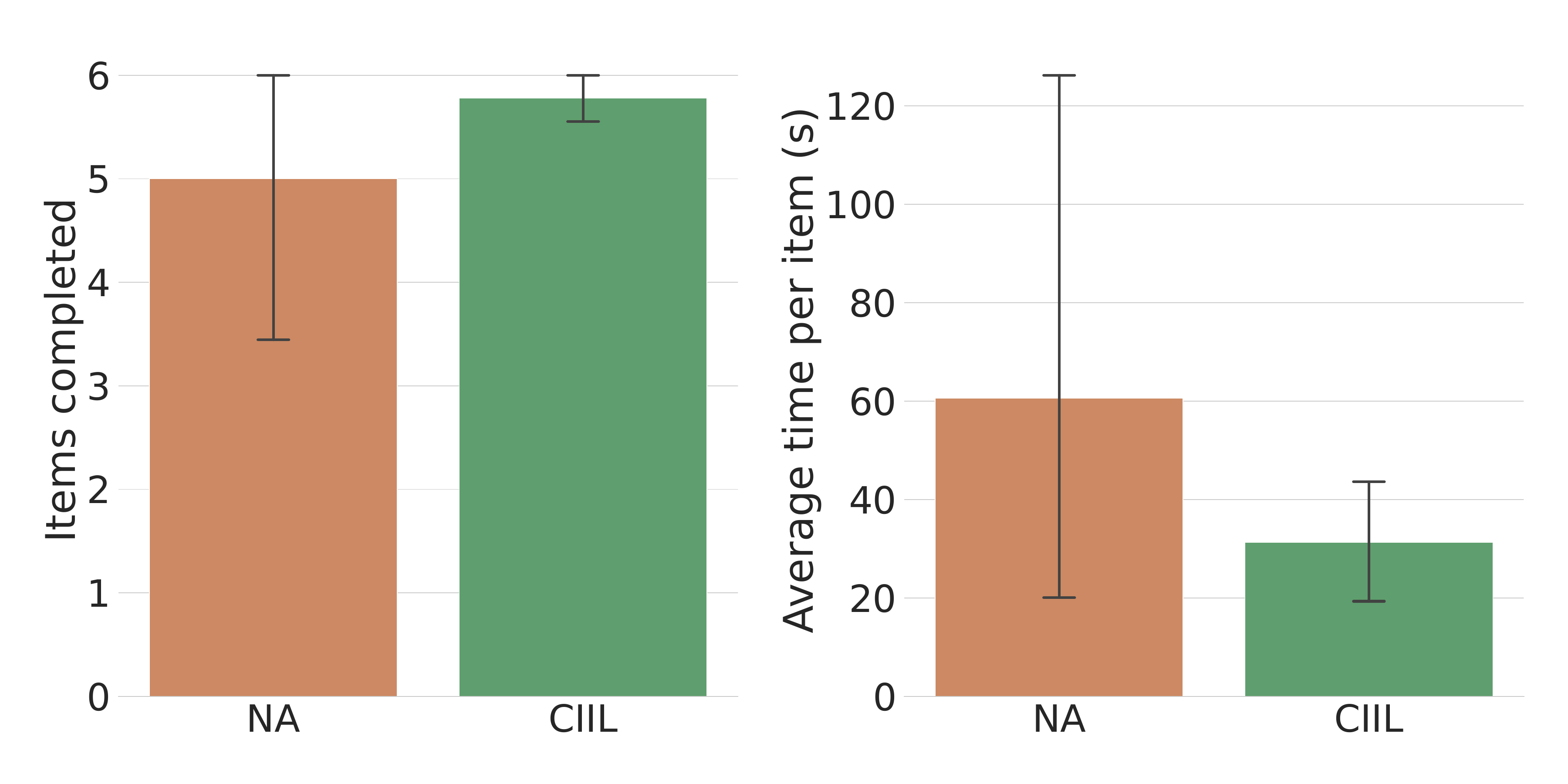}}
    \caption{Online myoelectric control performance metrics.}
    \label{fig:barplot-online}
\end{figure}

In the NA trials, participants completed $5.00 \pm 1.89$ objects on average with an average time taken for each object of $60.59 \pm 87.31$~s. As for the CIIL trial, $5.78 \pm 0.42$ objects were completed on average while the average time taken per object was $31.26 \pm 18.51$~s. These results are presented in Fig.~\ref{fig:barplot-online}. The NASA-TLX results are presented in Table~\ref{tab:tlx}, with the average of all dimensions in the last row. The NASA-TLX survey evaluates the perceived workload on a scale of 1-20, a higher value representing a higher perceived workload.

\begin{table}
    \centering
    \caption{NASA-TLX subjective workload assessment results}
    \begin{tabular}{l c c} \hline
    \textbf{Dimension} & \textbf{NA} & \textbf{CIIL}\\ \hline
    Mental & $7.83 \pm 7.24$ & $7.78 \pm 5.04$ \\
    Physical & $8.94 \pm 6.33$ & $5.61 \pm 3.79$ \\
    Temporal & $7.44 \pm 7.67$ & $5.11 \pm 5.18$ \\
    Performance & $11.17 \pm 6.22$ & $8.39 \pm 4.53$ \\
    Effort & $9.78 \pm 5.09$ & $9.61 \pm 5.73$ \\
    Frustration & $8.67 \pm 7.21$ & $8.78 \pm 7.14$ \\
    \textit{Overall} & $8.97 \pm 6.63$ & $7.55 \pm 5.23$ \\
    \hline
    \end{tabular}
    \label{tab:tlx}
\end{table}


\section{Discussion}
\label{sec:discussion}

This study evaluates the usability of self-supervised myoelectric control systems for goal-oriented tasks. Immersed in a VR environment, participants handled objects with the help of EMG. In addition to conventional offline and online metrics, this study also considered participants' subjective perception to better capture the usability of the control systems, addressing a gap often overlooked in prior works ~\cite{resnik2021structural}.

The Post-NA trial showed better accuracy than the Initial evaluation (Fig.~\ref{fig:results-boxplot}). This improvement is likely due to participants refining their gestures during the task, aided by the live and visual feedback provided in the VR environment. Previous studies confirm that visual feedback helps new EMG users improve muscle contraction quality, leading to more consistent muscle activation patterns~\cite{kim_semg-based_2022}.

The CIIL trial yielded a noticeable reduction in perceived workload compared with the NA trial, despite a lower offline accuracy. A potential explanation lies in the dynamic nature of CIIL: instead of requiring users to align their muscle contractions with fixed classification parameters as is necessary in the NA trial, the CIIL framework allows the classifier to evolve alongside the user's natural gesture patterns, adapting to the muscle contractions they find most intuitive and thereby bridging the gap between SGT-generated EMG and goal-oriented behavior. This might also explain how participants were able to complete more object transfers in less time and with less effort. In the CIIL trial, the adaptive classifier gradually improved control for these initially problematic gestures, whereas in the NA trial, users had to rely on trial-and-error to discover muscle activation patterns that worked, increasing both effort and frustration.

Overall, these findings highlight the benefits of incorporating adaptation into myoelectric control systems. Although the CIIL-adapted classifier exhibited lower offline accuracy, participants reported higher usability and achieved faster task completion, further highlighting the limited utility of offline accuracy as a sole predictor of real-world performance~\cite{campbell_screen_2024}.

While CIIL improved usability and reduced perceived workload, its effectiveness in optimizing efficiency and objective performance requires further refinement. The current relabeling strategy, which relies on label spreading for negative context samples, risks gradually eroding less prevalent classes if the adaptation dataset is imbalanced, potentially causing the system to ``forget`` these classes. This is seen in Fig.\ref{fig:results-boxplot}, where the accuracy of the HR gesture dropped from 91\% in the Initial test to 84\% after the CIIL trial. Addressing this limitation by exploring alternative smoothing strategies, such as balanced sampling or regularization techniques, could further enhance the robustness and practicality of CIIL EMG systems. Furthermore, long-term, cross-session studies are needed to verify the stability of CIIL-learned models across sessions. Also, leveraging multimodality could further improve CIIL. For example, IMUs~\cite{li_enhancing_2025} could provide the classifier with information about the arm's orientation, one of the confounding factors of EMG~\cite{campbell_current_2020}.

\section{Conclusion}

This study demonstrates the potential of CIIL to enhance myoelectric control systems by enabling real-time adaptation through contextual information. Tested in a realistic VR environment with a goal-oriented task, CIIL reduced perceived user effort and improved usability, particularly for participants with low baseline accuracy. CIIL enhanced system adaptability in dynamic scenarios, bridging the gap between lab settings and real-world applications. To further improve robustness, future research could explore alternative pseudo-labeling methods to mitigate the risk of imbalances in the adaptation dataset.

\bibliographystyle{IEEEtran}
\bibliography{main}

\end{document}